\documentclass[twocolumn,prl,showpacs,preprintnumbers,amsmath,amssymb]{revtex4}

\usepackage{graphicx}

\newcommand{\mm}[1]{\( #1 \)} 
\newcommand{\sog}[1]{\left( #1 \right)} 
\newcommand{\wayback}{\ \ \ \ \ } 

\begin{document}

\title{Fourier processing of quantum light}

\author{E. Poem}
\email{eilon.poem@weizmann.ac.il}
\author{Y. Gilead}
\email{yehonatan.gilead@weizmann.ac.il}
\author{Y. Lahini}
\author{Y. Silberberg}

\affiliation{
Department of Physics of Complex Systems, Weizmann Institute of
Science, Rehovot 76100, Israel }

\date{\today}

\begin{abstract}
It is shown that a classical optical Fourier processor can be used for the shaping of quantum correlations between two or more photons, and the class of Fourier masks applicable in the multi-photon Fourier space is identified. This concept is experimentally demonstrated using two types of periodic phase masks illuminated with path-entangled photon pairs, a highly non-classical state of light.
Applied first were sinusoidal phase masks, emulating two-particle quantum walk on a periodic lattice, yielding intricate correlation patterns with various spatial bunching and anti-bunching effects depending on the initial state. Then, a periodic Zernike-like filter was applied on top of the sinusoidal phase masks. Using this filter, phase information lost in the original correlation measurements was retrieved.
\end{abstract}


\maketitle

\begin{figure}[tbh]
\centering
\includegraphics[width=0.4\textwidth]{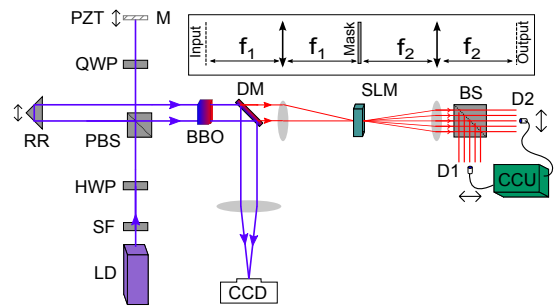}
\caption{Experimental setup. Polarized light from a laser diode (LD) is split into two beams by a half-wave plate (HWP) and a polarizing beam splitter (PBS). The distance between the beams is controlled by a movable retroreflector (RR), and their relative phase is controlled by a movable mirror (M) on a piezoelectric transducer (PZT). The beams undergo down-conversion in a nonlinear crystal (BBO). The pump beams are diverted by a dichroic mirror (DM) and interfered on a camera (CCD). The interference pattern is used together with the PZT for actively stabilizing the relative phase between the beams. The down-converted light goes through a 4-f filter equipped with a spatial light modulator (SLM), and finally detected by two fiber-coupled detectors (D1, D2). Coincidence detection is manifested by a coincidence counting unit (CCU). SF is a spatial filter, QWP is a quarter-wave plate, and BS is a beam-splitter. Purple lines represent the pump beams, and red lines represent the down-converted light. The inset shows a scheme of a 4-f filter.
} \label{fig_setup}
\end{figure}

Fourier processing is a well-established method in signal processing, where a transformation is applied to the signal in the reciprocal Fourier domain. In classical optics, a lens converts an optical field in the back focal plane to its spatial Fourier transform at the front focal plane, where optical operations can be applied to directly manipulate the signal in the fourier domain~\cite{book_goodman}. This basic scheme has been used for the control and manipulation of the light intensity distribution in a number of optical signal processing methods, from matched filtering to phase-contrast microscopy~\cite{book_goodman}. One of the most frequently used setups is that of the 4-f filter, where a spatial mask at the common focal plane of two lenses performs the Fourier processing of the input signal (see inset in Fig.~\ref{fig_setup}).

In recent years, there is a growing interest in the generation, control and detection of non-classical, quantum light~\cite{book_gerry_knight}. This is mostly due to its exciting applications in quantum information processing~\cite{review_quant_interf_pan}, quantum metrology~\cite{review_quant_metrology_lloyd}, quantum imaging~\cite{review_quant_imaging_shih}, and quantum lithography~\cite{review_quant_lithography_dowling}.
In contrast to classical light, many properties of quantum light cannot be revealed through first-order correlations, such as the light intensity, and require the measurement of higher order correlations~\cite{review_quant_interf_pan,book_ou}. Since optical Fourier processing proved to be a powerful tool for controlling the intensity distribution of classical light,
examining the possibility of using this technique for the control of quantum correlations between photons seems very promising.
This approach is further motivated by recent works applying Fourier optics to two photon states~\cite{art_Fourier_quantum_interferometry_th,art_Fourier_quantum_interferometry_exp,art_two_photon_wolf_equations_sergienko,art_fourier_lithography,art_fourier_angular_padjett,art_angular_two_qubit_padjett}, where it was shown that the optical Fourier transform of two photons in one spatial dimension is given by the two-dimensional Fourier transform of the two-photon wavefunction~\cite{art_two_photon_wolf_equations_sergienko,art_fourier_lithography}.

In this work,
we first establish an analogy between the Fourier processing of classical light (intensity) to that of quantum light (correlations) by studying the effect of the Fourier-plane mask on photon correlations. We show that the effect of a one-dimensional mask in the space of the two-photon wavefunction is equivalent to the two-dimensional, external product of the one-dimensional mask with itself. This defines the class of Fourier filters that can be applied in the two-photon correlation space using a classical, one dimensional mask. We then use these results to experimentally demonstrate the shaping of non-classical correlations between two photons by spatial Fourier processing with various masks.

For the experimental demonstration, path-entangled photon pairs, a highly non-classical state of light, were introduced into a 4-f filter. The resulting two-photon correlation maps were measured for two different types of phase masks implemented at the Fourier plane by a spatial light modulator (SLM). The first mask extended the correlation object by effectively implementing quantum walk of two particles~\cite{art_perets_waveguides,art_bromberg_correlations,art_obrien_correlations}, and the second mask was used as a Zernike-like filter, similar to that used in classical phase-contrast microscopy, for retrieving two-photon phase information.

Before we discuss the details of the experiments and the experimental results, we would first like to review the application of the optical Fourier transform to more than one photon, and use it to establish the analogy between single-photon and multi-photon Fourier processing.
For the simplicity of the following discussion we consider monochromatic light, omit the polarization degree of freedom, and consider only polarization maintaining, linear optical devices. A single-photon transition operator describing such a device has the form, $\hat{U}$=$\int dx_odx_iU(x_o,x_i)\hat{b}^{\dagger}(x_o)\hat{a}(x_i)$.
The operator $\hat{a}(x_i)$ [$\hat{b}^{\dagger}(x_o)$] annihilates [creates] a photon at the coordinate $x_i$ [$x_o$] of the input [output] plane of the device, and $U(x_o,x_i)$ is the (complex) amplitude of that transition. The coordinates can be either one- or two-dimensional, depending on the studied system, and the operators obey bosonic commutation relations.
We then introduce at the input plane of the optical device a two-photon state,
\begin{equation}\label{eq:Psi_in}
\Psi_{in}=\int dx_{i1}dx_{i2}A_S(x_{i1},x_{i2}) \hat{a}^{\dagger}(x_{i2})\hat{a}^{\dagger}(x_{i1})|0\rangle,
\end{equation}
where $A_S(x_{i1},x_{i2})$=$A_S(x_{i2},x_{i1})$ is an exchange-symmetric complex probability amplitude. The spatial intensity-correlation function of this state is given by $\Gamma(x_{i1},x_{i2})$=$|A_S(x_{i1},x_{i2})|^2$.
For two non-interacting photons passing together through the device, the output two-photon state is obtained from the input state by applying the single-photon transition operator twice, $\Psi_{out}$=$\hat{U}^2\Psi_{in}$. One then obtains,
\begin{equation}\label{eq:Psi_out}
\Psi_{out}=\int dx_{o1}dx_{o2}B_S(x_{o1},x_{o2}) \hat{b}^{\dagger}(x_{o2})\hat{b}^{\dagger}(x_{o1})|0\rangle,
\end{equation}
where the output complex probability amplitude, $B_S(x_{o1},x_{o2})$=$B_S(x_{o2},x_{o1})$, is given by,
\begin{eqnarray}\label{eq:B_S}
B_S(x_{o1},x_{o2})=\int dx_{i1}dx_{i2}A_S(x_{i1},x_{i2})\cdot\wayback\wayback \\
\cdot U(x_{o2},x_{i2})U(x_{o1},x_{i1}).\nonumber
\end{eqnarray}
The two-photon transition amplitude is thus just the external product, $U(x_{o2},x_{i2})U(x_{o1},x_{i1})$, of the single-photon transition amplitude with itself~\cite{art_two_photon_wolf_equations_sergienko}. In the same manner it can be shown that this property holds for an arbitrary number of non-interacting photons.

We note that this result, though derived here for photons, which are bosons, applies also to fermions, providing that an exchange-\textit{anti-}symmetric initial probability amplitude is being used.

We now apply this general result to the components of the Fourier processor. We begin by treating the lenses, which perform the Fourier transforms from the input to the Fourier plane, and from there to the output. Eqs.~(\ref{eq:Psi_out})-(\ref{eq:B_S}) applied to a one dimensional Fourier transform, like that performed, for example, by the first lens, $U_{L1}(x_f,x_i)$=$e^{-2\pi ix_fx_i/\lambda f_1}$, recover the known result that in the two photon coordinate space this transformation is equivalent to a two-dimensional Fourier transform, $U_{L1}(x_{f1},x_{i1})U_{L1}(x_{f2},x_{i2})$=$e^{-2\pi i(x_{f1}x_{i1}+x_{f2}x_{i2})/\lambda f_1}$~\cite{art_two_photon_wolf_equations_sergienko,art_fourier_lithography}. Here $x_f$ is the coordinate in the Fourier plane, $f_1$ is the focal length of the lens, and $\lambda$ is the wavelength of the incident light. This result applies to the second lens as well. It can also be generalized to any number of photons, N, and spatial dimensions, D, since the external product of N D-dimensional Fourier transforms is just the N$\times$D-dimensional Fourier transform.

The second component in Fourier processing is the mask implemented at the Fourier plane of the 4-f filter. Since the action of a mask with complex amplitude transmission $M(x_f)$ on a single-photon probability amplitude is  $A(x_f)\rightarrow M(x_f)A(x_f)$, from Eq.~(\ref{eq:B_S}) it follows that its action on a two-photon probability amplitude is $A_S(x_{f1},x_{f2})\rightarrow M(x_{f1})M(x_{f2})A_S(x_{f1},x_{f2})$. This represents a mask in the two-dimensional Fourier plane, $\{x_{f1},x_{f2}\}$, which is the external product of the one-dimensional mask $M(x_f)$ with itself. This result reflects the fact that a single linear system can act only locally on the photons, and induces the same transformation to each photon. Since unlike the two-dimensional Fourier transform, a general two-dimensional mask is not necessarily a product of a one-dimensional transformation by itself, only this special class of two-photon masks can be implemented using a classical, linear mask. The mask, therefore, is the element that limits the ways in which a classical, linear Fourier processor can process a non-classical two-photon wave-function. The implementation of a general mask would require the two photons to interact on the mask. This is not yet feasible with current technology, though progress is being made in this direction~\cite{art_zeilinger_nlqc}. Nevertheless, even with linear masks, several important processing algorithms can still be implemented.
For example, a Fourier mask transmitting only in a small aperture will block fermion pairs, while boson pairs will go through. This is since such a mask is equivalent to a small square aperture on the main diagonal of the two-particle Fourier space, and thus transmits only the zero order with respect to particle exchange. In contrast to boson pairs, fermion pairs, which wavefunction is exchange-anti-symmetric, do not have such a zero order.

For the experimental demonstration of two-photon Fourier processing we use phase-only masks.
In contrast to the aperture mask discussed above, phase-only masks have, in principle, no loss. On the other hand, they cannot change the total two-photon flux. They can, however, controllably redistribute it among different output two-photon configurations, as we demonstrate below.

\begin{figure}[tbh]
\centering
\includegraphics[width=0.48\textwidth]{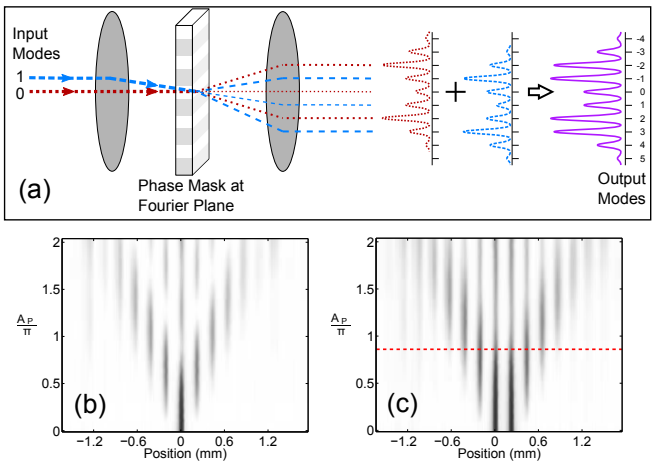}
\caption{(a) Schematic of the beam shaping system with a sinusoidal phase mask applied on the SLM, showing beam propagation and resulting intensity patterns. (b)-(c) Measured output intensity patterns for different values of the phase amplitude, $A_p$, for, (b), a single input beam, and (c), two beams of path-entangled photon pairs.}
\label{fig_modes}
\end{figure}

The experimental setup is described in Fig.~\ref{fig_setup}.
We first generate a two-path-entangled, two-photon state of the form, $\frac{1}{\sqrt{2}}(|2_a,0_b\rangle$+$e^{i\phi}|0_a,2_b\rangle)$, that is, $A_S(x_{i1},x_{i2})$=$\frac{1}{\sqrt{2}}[p(x_{i1}$-$x_a)p(x_{i2}$-$x_a)$+$e^{i\phi}p(x_{i1}$-$x_b)p(x_{i2}$-$x_b)]$, where $x_a$ ($x_b$) is the central lateral position of the first (second) path, and $p(x_i)$ is its lateral profile. This is achieved by down-conversion of a split, phase-stabilized laser beam~\cite{art_quantum_multimode}.
The down converted light then goes through a 4-f filter equipped with a one-dimensional, phase-only SLM at its Fourier plane. Using the SLM, various one-dimensional phase masks can be created. In this work we use only periodic phase patterns, and set their period to match the separation between the two initial modes, such that the output intensity pattern is also in the form of discrete modes [see Fig.~\ref{fig_modes}(a)]. Finally, the total output intensity distribution, as well as the output intensity correlations between any two modes $q$ and $r$, $\Gamma_{q,r}$, are recorded by two fiber-coupled single-photon detectors. Here \mbox{$\Gamma_{q,r}$=$\int_{-w}^{w}\int_{-w}^{w}dx_{o1}dx_{o2}|B_S(x_{o1}$-$x_q,x_{o2}$-$x_r)|^2$}, where the mode $q$ ($r$) is centered about $x_q$ ($x_r$), and $2w$ is the width of the collecting fibers.

The intensity distribution of the initial state is spatially limited to two spots. This is true also for the corresponding correlation pattern, since due to the bunched nature of the photons in the initial state, the correlation pattern is composed of two spots on the main diagonal. In order to demonstrate correlation manipulation similar to Fourier image processing, we first use a 4-f filter with a sinusoidal phase mask, $M\sog{x_f}$=$e^{iA_p\cos\sog{2\pi \nu x_f}}$,
where \mm{A_p} is the phase mask amplitude, and \mm{\nu} is the spatial frequency of the modulation.
This filter implements the tight-binding model~\cite{book_yariv,book_goodman}, and emulates the quantum walk~\cite{art_perets_waveguides} of the two entangled photons on a periodic lattice. This is since the phase mask is applied in the Fourier space -- the space of the eigen-modes of any periodic system, and since the induced phase shift is proportional to the sinusoidal tight-binding dispersion relation~\cite{remark_tight_binding}. This filter thus results in spread-out intensity distributions~\cite{book_yariv,book_goodman,art_perets_waveguides}, and intricate correlation patterns that depend on the phase, $\phi$, between the two initial modes~\cite{art_bromberg_correlations}.

Indeed, the measured output intensity patterns for a single input beam [two input beams], shown in Fig.~\ref{fig_modes}(b)[(c)], are practically identical to those observed in other optical systems implementing the tight-binding model, such as a periodic lattice of evanescently coupled waveguides~\cite{art_perets_waveguides,art_bromberg_correlations,art_obrien_correlations}. The incident light, originally concentrated in one or two input modes, spreads among a number of output modes increasing linearly with $A_p$.
\begin{figure}[tbh]
\centering
\includegraphics[width
=0.48\textwidth]{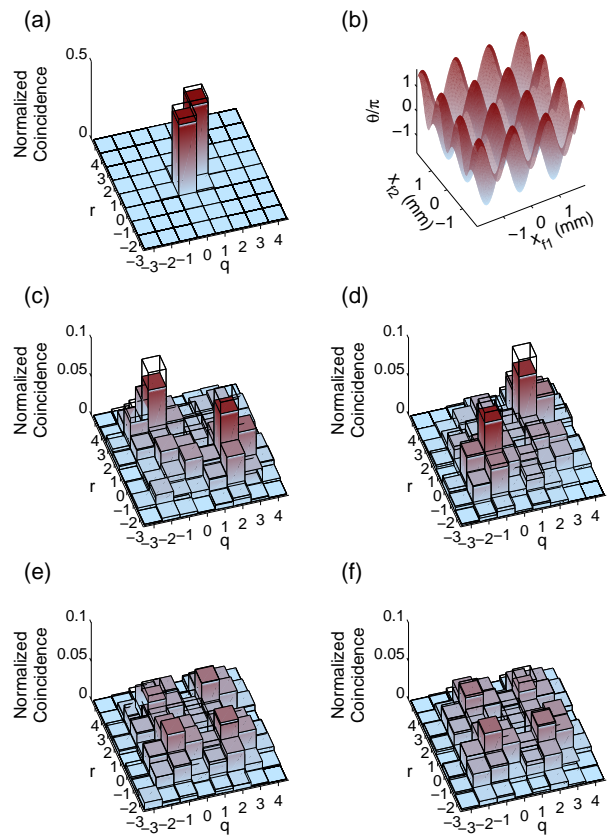}
\caption{ Experimental observation of non-classical correlations using a sinusoidal phase mask. (a) Second-order spatial correlation function, ${\Gamma_{q,r}}$, of the input state, as measured at the output when no mask was applied. The black frames represent the expected values. (b) A section of the phase pattern created by the applied sinusoidal phase mask in the two-photon Fourier plane. (c)-(f) Resulting correlation patterns for different phases, $\phi$, of the input states. Anti-bunching is observed for $\phi=0$ (c), while for $\phi=\pi$ (d) bunching is observed. For the two intermediate cases, $\phi=-\pi/2$ (e), and $\phi=\pi/2$ (f), both effects are seen simultaneously.} \label{fig_sine_phase_corrs}
\end{figure}

For the correlation measurements we apply a sinusoidal phase mask with $A_p=0.86\pi$. The corresponding intensity pattern is indicated by the dashed line in Fig.~\ref{fig_modes}(c). The phase of the two-dimensional mask that results in the two-photon Fourier space is presented in Fig.~\ref{fig_sine_phase_corrs}(b). The resulting correlation maps are presented in Figs.~\ref{fig_sine_phase_corrs}(c)-\ref{fig_sine_phase_corrs}(f), for four different initial phases, $\phi$. The full bars present the measurement, while the empty bars present the theoretical prediction. For comparison, the correlation map of the initial state, as measured at the output when no mask was applied, is presented in Fig.~\ref{fig_sine_phase_corrs}(a).

It is seen that, depending on $\phi$, the two photons, which are initially bunched together, can emerge anti-bunched [$\phi$=$0$, Fig.~\ref{fig_sine_phase_corrs}(c)], bunched [$\phi$=$\pi$, Fig.~\ref{fig_sine_phase_corrs}(d)], or in any super-position of these two extremes. Of these super-positions, those generated for the two cases of $\phi$=$-\pi/2$ and $\phi$=$\pi/2$ have identical correlation maps [See Figs.~\ref{fig_sine_phase_corrs}(e) and \ref{fig_sine_phase_corrs}(f)]. However, since the initial phase is different between the two cases, so should be the phases in the final patterns, and we would therefore like to have a procedure that will reveal this hidden information.
\begin{figure}[t]
\centering
\includegraphics[width=0.48\textwidth]{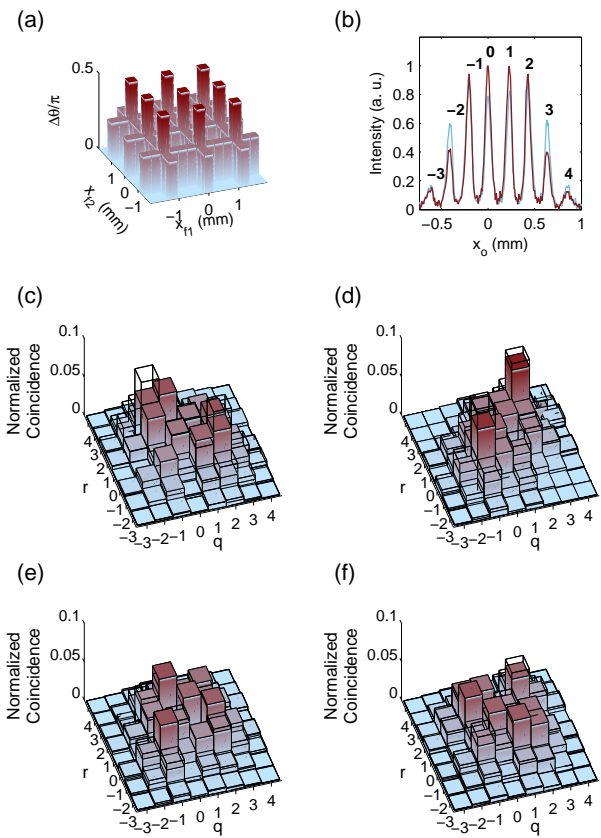}
\caption{Retrival of phase information using a Zernike-like filter. (a) A section of the additional phase pattern applied on top of the sinusoidal phase [Fig.~\ref{fig_sine_phase_corrs}(b)]. (b) A comparison between the output intensity distribution with (dark red) and without (light blue) the additional phase. The number of each peak is indicated next to it. (c)-(f) Resulting correlation patterns for the same initial phases as in Fig.~\ref{fig_sine_phase_corrs}(c)-(f). While for $\phi$=$0$ (c) and $\phi$=$\pi$ (d) no qualitative change is seen, the two intermediate states, $\phi$=$-\pi/2$ (e) and $\phi$=$\pi/2$ (f) now show different and opposite correlation patterns, revealing the phase difference between them.} \label{fig_comb_phase_corrs}
\end{figure}

The second Fourier processing algorithm, which will be performed on the obtained correlation patterns, would thus be dedicated to this purpose. We therefore add on top of the sinusoidal phase grating a second phase grating, implementing an algorithm designated to reveal the phase information lost in the correlation measurements discussed above. We note that this is equivalent of building a second 4-f filter and adding the second grating at its Fourier plane, since then the two Fourier planes would just be imaged one on top of the other.

The specific grating we use for this purpose is inspired by phase-contrast microscopy. In this method, the zero-order of the image is phase-shifted by $\pi/2$ with respect to all other orders using a designated Fourier phase filter, known as the Zernike filter. For a transparent, apparently uniform object, with a small spatial phase-shift distribution, this leads to interference of the non-zero orders with the zero-order, and the creation of intensity fluctuations proportional to the local phase-shifts~\cite{book_goodman}. For a periodic system, the zero-order is at the center of each reciprocal lattice unit cell. In order to approximate the Zernike filter for the case of two photons, we applied a $\pi/4$ phase to the central quarter of each phase lattice unit cell (with respect to the original sinusoidal phase). In the two-dimensional space of the two photon wavefunction this filter shifts the central part of each two-dimensional unit cell by $\pi/2$. There is, however, a byproduct: a $\pi/4$ shift along the main axes of each cell. The filter phase is shown in Fig.~\ref{fig_comb_phase_corrs}(a). Our filter thus only approximates the Zernike filter. Furthermore, the initial correlation patters [Figs.~\ref{fig_sine_phase_corrs}(e) and \ref{fig_sine_phase_corrs}(f)] are non-uniform, and the predicted phase shifts are not small. Nevertheless, the filter can still recover qualitative phase information, and differentiate between the two cases.

This is shown in Fig.~\ref{fig_comb_phase_corrs}. First note that, as seen in Fig.~\ref{fig_comb_phase_corrs}(b), the filter has almost no influence on the intensity distribution. This is due to its unitarity and its periodicity, matching that of the sinusoidal phase mask. The filter also does not qualitatively change the correlation patterns for the cases of $\phi$=$0$ [Fig.~\ref{fig_comb_phase_corrs}(c)] and $\phi$=$\pi$ [Fig.~\ref{fig_comb_phase_corrs}(d)]. This is since the wavefunctions in these cases are all real, and thus the Zernike-like filter has little or no influence on them.
In sharp contrast, for the cases of $\phi$=$-\pi/2$ [Fig.~\ref{fig_comb_phase_corrs}(e)] and $\phi$=$\pi/2$ [Fig.~\ref{fig_comb_phase_corrs}(f)], the application of the Zernike-like filter changes the correlation patterns, and creates a clear difference between these two cases, which now show opposite, asymmetric correlation patterns.

In conclusion, we have demonstrated that the correlations between path-entangled, two photon states can be manipulated using a classical Fourier processor. Though the applicable masks are limited to local operations, many types of manipulations are still possible. We used a phase-only SLM, and applied phase masks both to create an elaborate correlation pattern through the emulation of quantum walk on a periodic lattice, and to recover hidden phase information. For doing so we used methods adapted from classical Fourier processing, such as phase contrast microscopy. The analogy between image and correlation Fourier processing can be further exploited by using filters that change the amplitude of the light as well as the phase. Examples of such filters include \mbox{low-,} \mbox{high-,} or band-pass filters, the schlieren filter, and matched filters. Furthermore, additional degrees of freedom can be utilized by considering two-dimensional and/or polarization changing masks. Moreover, this analogy is applicable also in the time-domain, where it can be applied for the tailoring of temporal correlations between photons through pulse-shaping techniques~\cite{art_temporal_shaping_of_entangled_photons}. We thus believe that the use of Fourier processing methods may open new avenues for the manipulation of non-classical light.

We would like to thank Yonatan Israel for his useful suggestions, and Ron Sabo and Yaron Bromberg for their contributions in the early stages of this work. The financial support of the Minerva
Foundation, the European Research Council, and the Crown Photonics Center is
gratefully acknowledged.


\begin{thebibliography}{21}
\bibitem[{rem()}]{remark_tight_binding}
\bibinfo{note}{Note that this is not limited to the tight-binding model, and
  can be straightforwardly generalized to any periodic system with a
  single-valued dispersion relation.}
\end{thebibliography}
\end{document}